# Mirror Effect from Atomic Force Microscopy Profiles Enables Tip Reconstruction

Francisco Marqués-Moros, Alicia Forment-Aliaga, Elena Pinilla-Cienfuegos, and Josep Canet-Ferrer*

[1] Instituto de ciencia molecular (ICMol), Universidad de Valencia, Paterna, Spain

E-mail: jose.canet-ferrer@uv.es

**Abstract**

In this work, the tip convolution effect in atomic force microscopy is revisited to illustrate the capabilities of cubic objects for determination of the tip shape and size. Using molecular-based cubic nanoparticles as a reference, a two-step tip reconstruction process has been developed. First, the tip-to-face angle is estimated by means of an analysis of the convolution error while the tip radius is extracted from the experimental profiles. The results obtained are in good agreement with specification of the tip supplier even though the experiments have been conducted using real distribution of nanoparticles with dispersion in size and aspect ratio. This demonstrates the reliability of our method and opens the door for a more accurate tip reconstruction by using nano-lithography patterns.

## 1. Introduction

Atomic force microscopy (AFM) is a powerful tool widely used for imaging surfaces with nanoscale resolution [1–3]. AFM makes use of a nanometric tip for scanning the surface of a given sample while mapping the tip-sample interaction to reproduce its morphology. This enables its application in fields as different as biotechnology [4], semiconductors [5], photonics [6], polymer science [7] or 2D materials [8].

One of the main limitations of the AFM is the lateral resolution which comes from the finite size of the probe tip [9–11]. When the tip and the scanned motifs are comparable in size, the width measurements are dramatically overestimated [12], and thus, the shape of the motifs cannot be accurately resolved [13,14]. This effect is the so-called Tip Convolution (TC), and the estimation of the real dimensions of the surface objects is called tip deconvolution [15–17]. As TC has got a clear dependence on the size and shape of the used probe [18] few tip reconsctruction techniques have been developed with the aim to allow more accurate tip deconvolution [1].

All the approaches for AFM tip reconstruction present pros and cons [19]. Direct approaches, like for example imaging the probe by electron microscopy [20,21], offer an accurate picture of the probe. However, these strategies are time consuming and require undesirable manipulation of the probes [22,23]. Other methods are based on the use of calibration nano-patterns comparable to the probe tip in size. The resulting images are assumed to contain equal amount of information about the sample and the tip [19], to be extracted numerically. However, the resolution is strongly dependent on the nanofabrication limitations [24]. Finally, we can find a range of Blind Tip Reconstruction (BTR) algorithms which have been developed to estimate the probe geometry without knowing the specific motifs on the sample [19]. As a disadvantage, BTR methods are based in mathematical morphology [25–27] and require a wide mathematical background to be understood.

In our previous work [13], we proposed a simple algorithm capable of determining convolution effects by means of the numerical simulation of the scan performed by an AFM tip. This simulation was proved to work for diverse nano-objects with different geometries. This way, we could illustrate the influence of the size, shape, height and aspect ratio of different nanometric motifs placed over a flat surface. We also demonstrated that, in most cases, the real width of a surface motif can be estimated from the acquired AFM images without knowing its geometry in detail.

As an extension of our previous work, here we provide a more detailed description of the tip-sample interaction. Based on this description, we can re-define the convolution operation and identify those points at the images containing information about the tip. We demonstrate that in the case of rectangular objects, the information of the tip is acquired due to an extended interaction at the corners of the object. As a result, the tip geometry is clearly reflected in the resulting AFM profile, with minor distortion. Finally, as experimental proof, tip reconstruction experiments are carried out using cubic nanoparticles (NPs).

## 2. Background and Experimental Approaches

### 2.1 Definition of the AFM lateral resolution in terms of mathematical convolution

As above mentioned, the lateral resolution in AFM images is limited by the contact area between the tip and the inspected object, which depends on diverse factors such as the size of the probe and the height of the object.

The effect of the TC has been treated from different points of view in literature [25,28–30]. As in the case of the BTR, most of the methods for description of TC are based on mathematical morphology or on geometrical considerations, where the tip and surface shapes are approached to analytical functions such as circumferences or parabolas [31–34]. Therefore, this limits the number of particular cases where the TC can be calculated by means of the proper mathematical operation of convolution.

To shed light on this, in Fig. 1 the result of convolution operator and the numerical simulations of AFM profiles are compared in opposite circumstances. In Fig. 1(a), the case of an AFM tip scanning a Dirac delta function is considered. As expected, the result consists of an inverted peak function reproducing the AFM tip profile. In this case, the analytical result of the convolution operator perfectly fits with the result of our numerical simulation software (not shown). In terms of AFM imaging, this means that the tip shape can be inspected by studying extremely narrow details. Nevertheless, not many nano-structures fulfil the condition of being extremely narrow, and unfortunately, slight deviations with respect to the delta-like function drive to an important reduction of the lateral resolution. For example, in Fig. 1(b) the result of the mathematical convolution between the tip and a step-like function (more similar to a real nano-structure), is plotted in dark yellow. The result is compared with the corresponding numerical simulation [Fig. 1(c)]. In this second case, the mathematical convolution does not fit with the numerical simulation. Indeed, the mathematical solution resembles neither the shape of the tip nor the step-like function. The reason of this mismatch will be treated later on, first, we will try to extract additional information from the numerical simulation.

The shape of the numerical profile in Fig. 1(c) could be understood as the broadening of the step-like function (due to the finite tip size) but also as an inaccurate replica of the AFM probe, where the tip has not been resolved by the step-like shape. Remember that AFM images are acquired by monitoring the interaction between the tip and the surface, hence, the image profiles contain the information of both geometries.

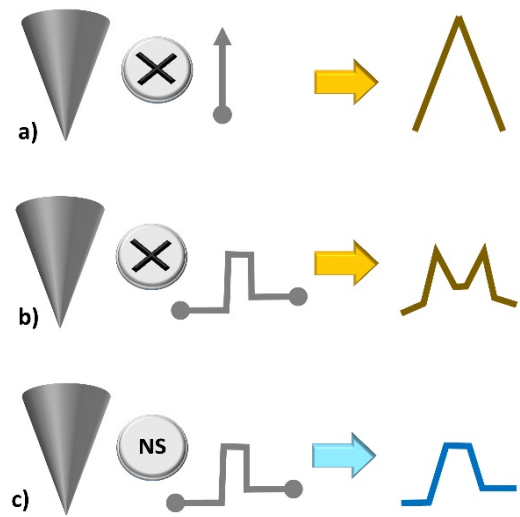

**Figure 1.** (a) Mathematical convolution between a tip and a delta function: the result fits the numerical simulation of the AFM profile. (b) Mathematical convolution between a tip and a step-like function and (c) the corresponding numerical simulation.

With this in mind, we can re-define the integration limits of the convolution operation to evaluate the step-like function profile at different stages of the scan, see Fig. 2. The first step is correlated with the scan from point O to A. During this step, the contact with the AFM probe is limited to the right side area of the tip, i.e. from T to R, Fig. 2(a). The second step begins with the tip reaching and scanning the top of the object, Fig. 2(b). Then, the contact is limited to the tip apex (point T) on the region A—B of the sample. The last step is depicted in Fig. 2(c), with the tip scanning the region B—C of the sample while making contact in the area comprised between L and T points (left side of the tip).

After evaluating the scan on separated sections, we found that the tip-sample interaction is restricted to certain areas in each step. Importantly, the convolution operator ($\otimes$) recovers the numerical simulation results if the operation is limited to the areas under interaction; i.e. T—R $\otimes$ O—A, T $\otimes$ A—B and L—T $\otimes$ B—C, respectively. The analytical result of these operations (blue solid lines in Fig. 2) is plotted on the profiles previously calculated using our software (light blue dashed lines in Fig. 2).

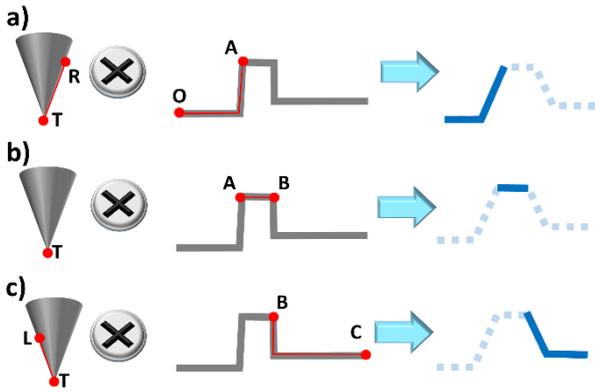

**Figure 2.** AFM tip operating on a step-like function analysed by steps. Three different interaction areas are defined: (a) from point O to A [R to T in the tip], from A to B [T in the tip], and (c) from B to C [T to L in the tip]. The analytical result (solid blue lines) recovers the numerical simulation (dashed clear blue lines) when limited to the corresponding interaction areas.

*2.2 Maximum expected error and "Mirror effect" in rectangular objects.*

Together with the tip size, the object height is the most influent parameter in the TC error. Considering objects of the same height, the larger TC is expected in the case of rectangular motifs, as demonstrated in Ref. [13]. The rectangular objects can be evaluated in three steps, as done with the step-like function. Again, the convolution is restricted to half a tip (T—R and L—T) at left and right top corners of the motif during the first and the third steps. In the course of the second step, only the tip apex interacts with the top side of the motif, section A—B. As a result, the second step of the profile is not affected by the tip size in this part of the scan. Indeed, the convolution error is associated to the first and the third steps where a wide area of the tip interacts with the corners of the object. At these points, the rectangular objects act as delta functions revealing the area of the tip where the contact has taken place[†].

Thanks to this, one half-side of the tip shape is reflected at each side of the AFM profile, without any distortion. This behaviour gives to rectangular objects the capability of imaging the shape of the tip, as a mirror does. This kind of "mirror effect" is illustrated in Fig. 3 by comparison of three numerical profiles. The numerical simulations have been carried out considering three different tip shapes, scanning a narrow [Fig 3(a)] and a wide rectangular object [Fig 3(b)]. Since both objects have the same height, they produce identical convolution error, i.e. the "tip reflection". As a result, the tip geometry can be obtained after removing the A—B section of the profile. For more details, a graphical reconstruction of those tips is illustrated Fig. SI1.

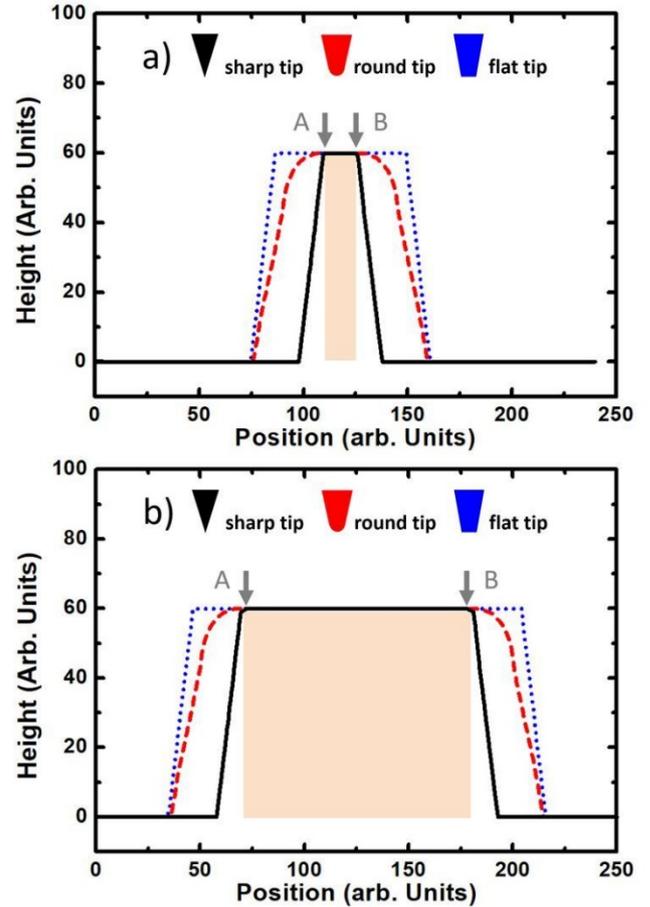

**Figure 3.** Numerical simulation of profiles from a narrow (a) and a broad (b) rectangular object (calculations were made in our previous paper [13]). Different tip shapes are considered: sharp (black line), round (red line) and flat (blue line). Because of the rectangular geometry, the same profile is obtained after removing section A—B. In addition, the remaining profile fits the corresponding inverted tip-shape giving place to the "mirror effect".

*2.3 Experimental methods: height to width correlation and averaged "mirror effect"*

---

[†] This is equivalent to employ the corners for imaging each half-side of the tip.

Based on this finding, here we propose two complementary approaches that can be combined to give rise to a novel tip reconstruction method.

The first approach consists on a statistical study of the experimental width measurements in comparison to the object height (i.e. h-w curve). This will be done by analysing the topography profiles of a large number of cubic NPs. This particular case of rectangular NPs allows settling a one-to-one correspondence between the real object width ($w$) and the height ($h$) measurements. Then, the convolution error can be accurately determined as follows [13]:

$$\tfrac{1}{2} w_{exp} = r_{tip} + \Delta + \tfrac{1}{2} w \quad (1)$$

where
$$\Delta = (h - r_{tip}) \tan(\gamma) \quad (2)$$

This way the relevant tip parameters $\gamma$ (tip-to-face angle) and $r_{tip}$ (tip radius) are bound to the real dimensions of the NPs, $h$ and $w$. Moreover, the experimental width ($w_{exp}$) is overestimated due to the TC error ($r_{tip} + \Delta$). Hence, Eq. 1 and Eq. 2 can be employed to estimate the tip parameters from the h-w plot.

The second approach consists on the reconstruction of the tip shape, as done with the numerical profiles in Fig. 3. This task results a bit more demanding when working with experimental profiles since they can be affected by surface tilt, roughness or shape irregularities. In the next section, we will show how the influence of these undesired effects can be identified and partially prevented by averaging a great number of profiles.

## 3. Results and Discussion

### 3.1 Cubic NPs as a reference

For this study we chose molecular-based nanoparticles of the bimetallic cyanide-bridged coordination networks known as Prussian blue analogues [35–37]. Specifically, we have used $K_{0.22}Ni[Cr(CN)_6]_{0.74}$ nanoparticles (KNiCr-NPs) that posses the face centred cubic structure (Fm3m) and grow as almost perfect cubic crystals an average size of 25 nm (23.9 ± 5.5 nm) and an average aspect ratio (AR) between 1 and 1.1. For further details about synthetic approach and physical characterization see our previous publications [38,39]. These NPs can be easily prepared in a large amount within a short time, by a straightforward solution method in water. They are obtained with a narrow size distribution and moreover, their size can be tuned at will by changing the alkaline cation [40]. Furthermore, thanks to their negative charge, they are directly stabilized in water as bare particles, without any organic capping. This last point is relevant for the correct performance of profile studies carried out for the development of this work because no additional adhesion forces between tip and organic capping are expected.

KNiCr-NPs were randomly deposited on a silicon substrate previously functionalized with a protonated amminopropyl-triethoxysilane (APTES) self-assembled monolayer. KNiCr-NPs solution was drop casted on top of the modified substrates which were copiously rinsed with mili-Q water after 1 minute to remove not attached NPs. Electrostatic driving forces between the cationic layer and the anionic nanoparticles drive the successful deposition of the particles [28].

### 3.2 w-h plot approach

Fig. 4(a) shows an AFM image of a representative area of the sample used for the tip characterization. From this kind of images, we can extract the height to width dispersion statistics, [Fig. 4(c)]. The dispersion suggests certain size anisotropy that we attribute to deviation of the NP dimensions with respect to the cubic geometry and to the acquisition error. In the case the NPs here studied (AR = 1.1) a deviation of about 10% in the measurement of the experimental width might be as we are considering AR=1. This deviation is perfectly acceptable for NPs about 25 nm in height, given the discretization of the AFM image is also around 10%, (i.e. 1x1 µm image with 512x512 pixel resolution).

According to the manufacture parameters the tip radius (PPP-NCH from *Nanosensors TM,* nominally $\gamma$ = 19-32º (depending on the tip orientation) and $r_{tip}$ = 5-10 nm) is clearly smaller than the NPs studied. The data analysis is carried out using just isolated NPs showing regular profile. Few representative profiles are shown in Fig. 4(b). Up to certain point we have included in our study NPs with slightly asymmetric profiles or moderate tilting [see black profile (v) in Fig. 4(b)]. These could be explained by a few angle tilt on the sample or by slight deviations from the rectangular geometry. Those defects are also contributing to dispersion observed in the *h-w* data.

The best fitting of Eq. 1 to the experimental data occurs for $\gamma = 19 \pm 3$º and the $r_{tip} = 15 \pm 6$ nm. We can conclude that the tip-to-face angle is successfully determined while the tip radius is clearly overestimated by means of this approach. The explanation is intrinsic to the w-h curve which allows to determine $\gamma$ directly from the slope. In contrast, $r_{tip}$ must be obtained by extrapolation, and hence, it can be dramatically affected by small changes in the slope of the curve. In this situation, we should look for better estimation of the tip radius by exploiting the "mirror effect".

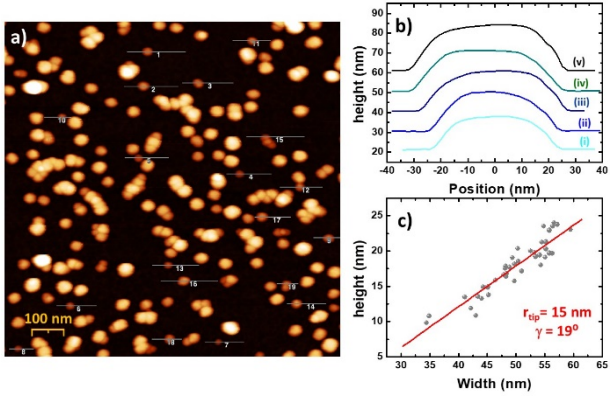

**Figure 4.** (a) AFM image of a representative area of the NPs sample under study. Few profiles can be measured in each image if focusing on isolated NPs with regular shape and avoiding aggregates. (b) Profiles of different height exhibiting a progressive experimental width reduction. (c) height-to-width plot generated with data from 46 NPs in three different images. The red line corresponds to the best fitting to Eq. 1.

*3.3 "Mirror effect" approach*

In this section the "mirror effect" is studied on the same profiles employed on the h-w curve. In Fig. 5(a) we show three examples: The top profile (in dark cyan) corresponds to a 20 nm height NP. At the upper side of this profile we clearly observe a plateau (marked in the figure with small black arrows), which would be related with the A—B section. This plateau is also observed in the numerical simulation of square objects [see red line in Fig. 2(b)]. The light blue profile (at the bottom), corresponds to a 15 nm height NP, and interestingly, it looks rather similar to the top one. In contrast, the A—B section cannot be easily distinguished in the slightly rounded profile at the centre, which corresponds to a 20 nm height NP (in blue). However, it still presents symmetry enough to be processed considering the maximum of the profile as the centre of the NP.

The results of removing the A—B section are shown in Fig. 5(b). We have used a simple sequential code of logic and arithmetic operations to identify and remove a segment of $w = h \pm 2$ nm (i.e. ± 1 pixel) width from each profile. This way, we also prevent a subjective assessment due to manual treatment on the data. Finally, the resulting profiles are inverted and biased with different offsets (the NP height) for a better comparison. The details of the procedure are offered in the Supporting Information. The processed data are shown as a diluted scatter plot in Fig. 5 (c). The broadening can be attributed to the rounding during the tip reconstruction data process. The red curve corresponds to the best fitting to the tip geometry which offers a good estimation of the relevant parameters despite the broadening: i) $r_{tip} = 7 \pm 1$ nm on both sides, close to the 5 nm estimated by the manufacturer; and ii) $\gamma = 20 \pm 8°$ also in good agreement but with an important relative error.

After comparing the result from different samples by using several tips, we can conclude that even the scatters in Fig. 5(c) clearly reproduce the tip shape and size, it must be taken as an upper bound since logical operations tend to produce important deviations in case of tilted profiles or defects (see Fig. SI2 for more details). Much better estimation can be obtained doing a manual treatment to minimize contribution of the artefacts and reduce the rounding error.

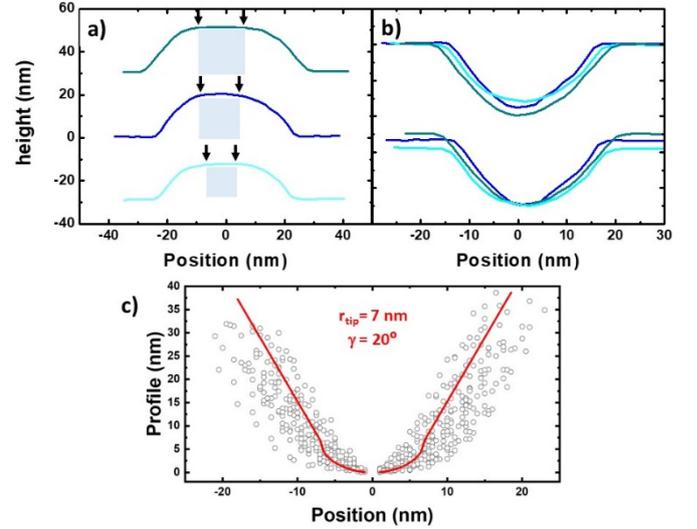

**Figure 5.** (a) Experimental profiles of three representative NPs. The arrows point the A—B section. (b) Reconstructed tip shape after removing the corresponding A—B section. For a better comparison the resulting profiles are fixed using both offsets: the sample surface and the tip apex. (c) Averaging of the 46 profiles extracted from three different images. The best fitting to Eq. 1 is plot in red.

Notice that our experiments have afforded the convolution in the most problematic range. In the case of NPs around 20 nm the height is too high for extracting information from the tip radius from w-h curves, but not enough for the proper estimation of $\gamma$ by means of the "mirror effect". The use of metrology patterns, instead of NPs, will definitely reduce the experimental data dispersion for both methods. However, we believe that using real samples results a more interesting case for the reader. Moreover, all the employed NPs can be easily synthetized in different sizes by a quick and inexpensive solution method. This represents the possibility of preparing multiple reference samples for the evaluation of different kind of tips by an approach that, in contrast to sophisticated lithography methods, is available to a broader number of research groups that commonly use AFM techniques in different fields. However, it is also worth mentioning that the above discussed sources of error in our results, could be reduced to the single pixel level by using lithographic patterns.

## 4. Conclusions

In this work, a deep discussion on the tip convolution effect is employed to illustrate the difficulties of analytical methods for estimation of convolution errors. For the sake of simplicity our discussion is focused on rectangular geometries which exhibit a prominent interaction with the tip at the corners of the object, leading to a high convolution effect. During this interaction the corners behave as delta-like functions, each one operating over a half-side of the tip, as we demonstrated by reproducing numerically an inverted image of the tip from simulated AFM profiles.

The experimental realization of our numerical predictions has been carried out using a dispersion of cubic molecular-based NPs as a reference sample for the tip reconstruction of a conventional AFM probe. We have evaluated the relation between the height and the width observed at the NP profiles, which relation has got analytical solution for rectangular shapes. This way, we can obtain a good estimation of the tip-to-face angle but a poor estimation of the tip radius. In a second approach, we have directly identified those points of the experimental profile containing the information of the tip ("mirror effect"). To ensure an objective assessment of tip reconstruction, in this step the experimental profiles have been treated by means of a sequential algorithm. This way we have found an accurate estimation of the tip radius. Therefore, both approaches are complementary and used for double checking the tip reconstruction.

## Acknowledgements

This work has been suported by the Spanish MINECO (Grants MAT2017-89993-R and Excellence Unit María de Maeztu MDM-2015-0538), and Generalitat Valenciana (PO FEDER, Reference IDIFEDER/2018/061). Financial support from the Spanish Ministry of Science co-financed by FEDER (Reference EQC2018-004888-P) is acknowledged. J. C.-F. thanks support from the Generalitat Valenciana by means of the program GenT (CIDEGENT/2018/005). A. F.-A. thanks the Universitat de València for her Senior Researcher contract. We thank Eva Tormos for technical support and helpful discussions.

# Supplementary Information

**Francisco Marqués-Moros, Alicia Forment-Aliaga, Elena Pinilla-Cienfuegos, and Josep Canet-Ferrer***

[1] Instituto de ciencia molecular (ICMol), Universidad de Valencia, Paterna, Spain
E-mail: jose.canet-ferrer@uv.es

***SI1. Graphical tip reconstruction from a narrow rectangular object.***

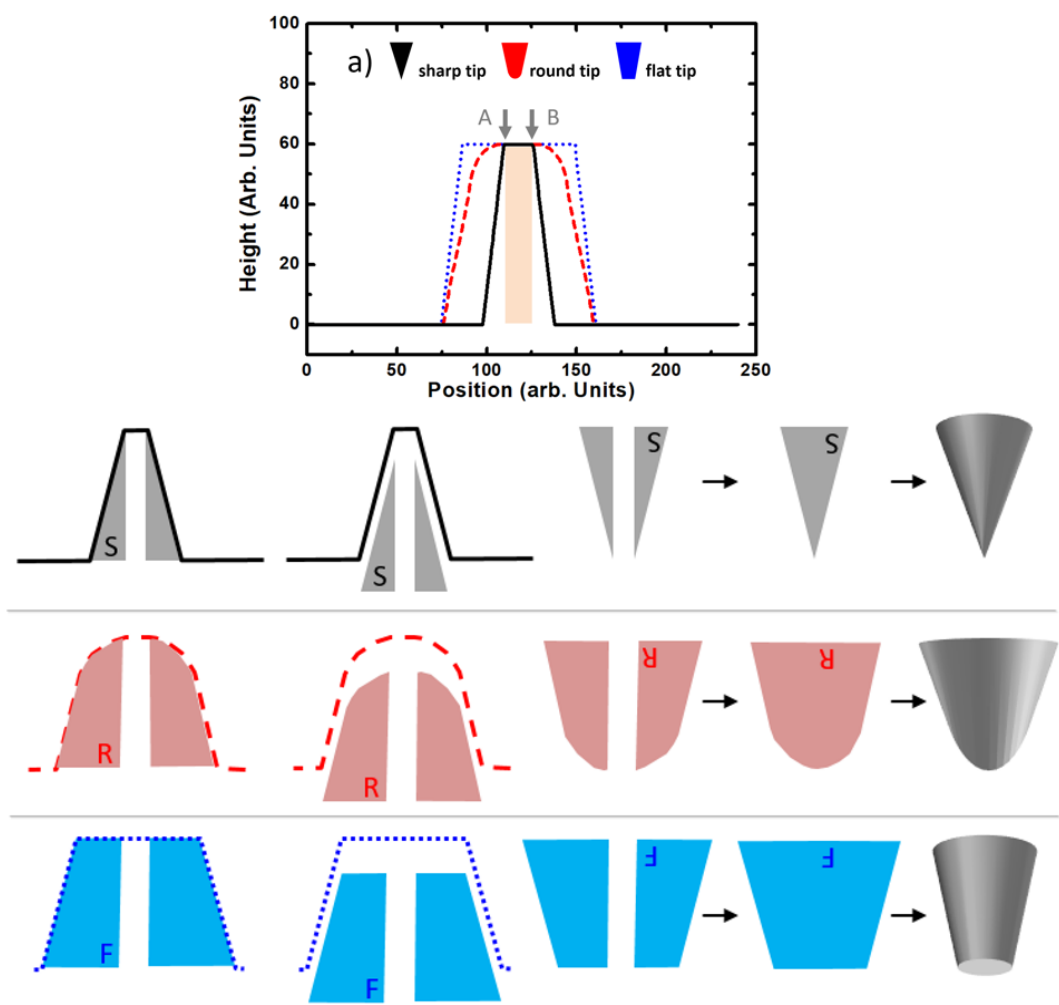

**Figure SI1.** Schematic representation of the tip reconstruction based on the mirror effect. It is presented step by step, showing how the corresponding tip-shape is obtained from he numerically simulated profiles (S for sharp tip; R for round tip; F for flat tip).

## SI2. Profiles processed in the manuscript and steps done for tip reconstruction.

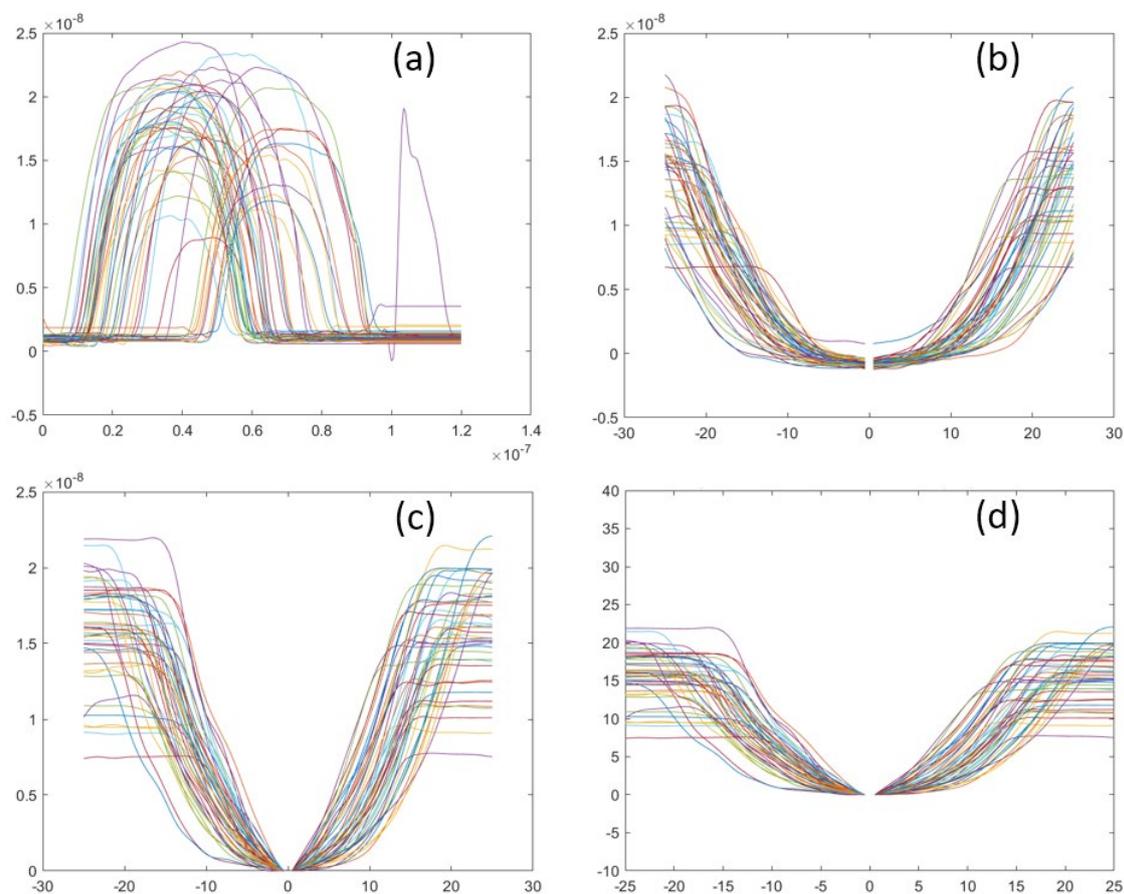

**Figure SI2.** Tip profile plots processed by means of a sequential algorithm in order to analyze numerically and average all the measured profiles. (a) The 46 profiles are interpolated to the same values in the abscissa (position) for an easier processing; then, they are inverted and aligned to the respective maxima (b). (c) The profiles after width correction –half-a-width of the height is removed in both sides–. (d) Rescaled profiles with units converted from meters to nanometers. Eventually, in these plots we can find some artifacts related with the acquisition of non-symmetric profiles (profiles with an odd number of data points) or sample tilt shifting the maxima from the center.

### *SI3. Algorithm to process the experimental profiles for tip reconstruction.*

In order to ensure an objective assessment of the tip reconstruction, the data are processed by means of the algorithm below.

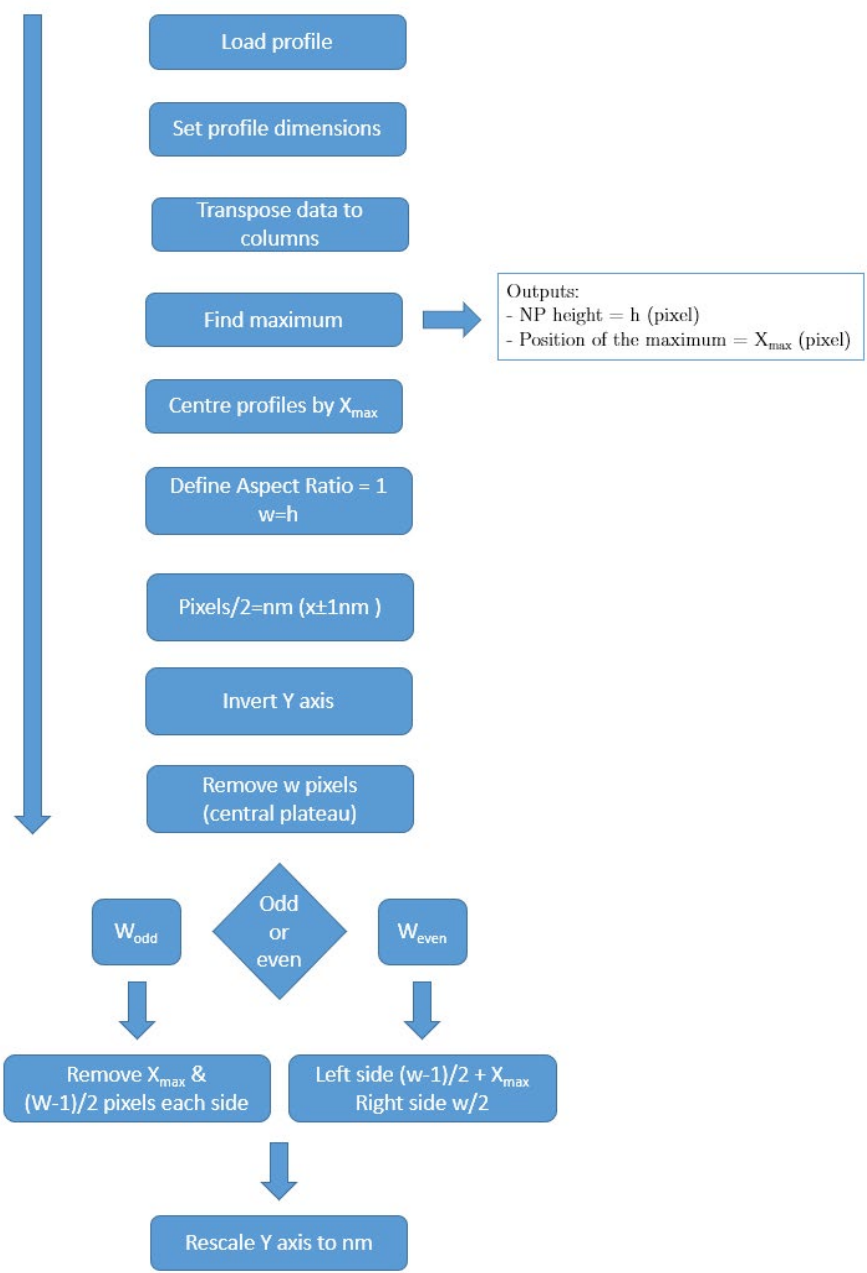

**Figure SI3.** The script uploads all the profiles as variables and it interpolates them for easing parallel processing, the unit of the profiles are pixels most of the time. Afterwards, the maximum of every profile is identified, so they can be ordered with respect to the maxima (considered as center of the profile). Then, the real width of the object is determined according to the aspect ratio to estimate the object width (w = h for cubic objects). After that, the script rounds the height (in pixels) to remove half-a-width pixels in both sides of the profiles. To conclude, the profiles are re-scaled and *y* variables are converted to nanometers.

## SI4. Artefacts at the experimental profiles.

When applying the h-w curve approach to a real case, the experimental tip shape profiles might present irregularities. For example, in the case cubic NPs we have observed several profiles exhibiting asymmetric shape, despite of the symmetry of NPs. These asymmetries can appear for several reasons such as sample tilt, scanning artefacts or defects on the NPs.

In Fig. SI4(a), we show an example of the proper tip reconstruction from a considerably symmetric profile. In Fig. SI4(b), we show a profile with a noticeable protrusion in the tip, which avoids an accurate reconstruction. In Fig. SI2(c), the maximum of the profile is not at the centre of the tip. In those cases, it is not possible to identify the proper tip shape. The accuracy of our results could be improved just by removing irregular profiles [like (b) and (c)] from the statistics, but we do include them to show the robustness of the method. From our results we can conclude that a certain error in height measurement is assumable. In the next section we will give some details.

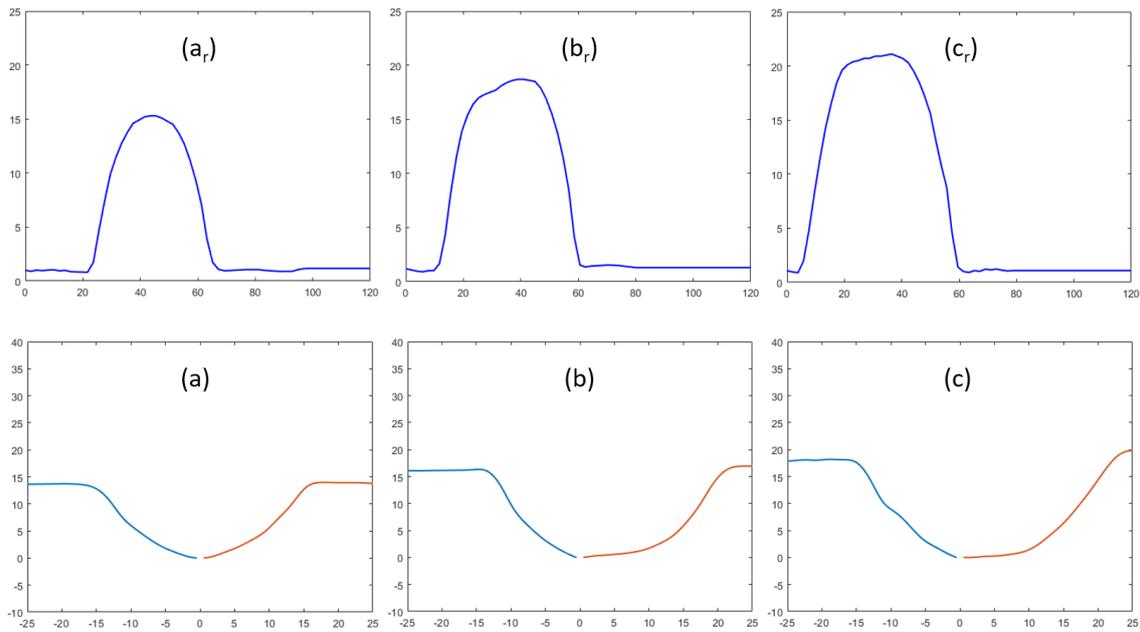

**Figure SI4.** Tip reconstruction for three different profiles used in the manuscript. ($a_r$) corresponds to the row data of a symmetric profile, but eventually we can find asymmetric profiles like ($b_r$, $c_r$). Those artifacts have a minor impact in the h-w curve, however they might affect the tip reconstruction, as shown in (b) and (c).

## SI5. Height-to-width (and width-to-height) fitting.

### SI5.1 Object larger than the tip radius.

As described in the manuscript, the convolution error can be estimated from the w-h relation for rectangular motifs larger than the tip radius, by using the expressions:

$$\frac{1}{2}w_{exp} = r_{tip} + \Delta + \frac{1}{2}w \tag{1}$$

$$\Delta = (h - r_{tip})\tan(\gamma) \tag{2}$$

being $w_{exp}$ the experimental width profile, $w$ the real width and $h$ the height of the object under inspection. The convolution effects are the sum of the tip radius ($r_{tip}$) and $\Delta$ which for objects larger than the tip radius accounts for the contribution of the tip-to-face angle ($\gamma$). Working with cubic objects we can consider $w=h$, and hence, Eq. (2) can be substituted in Eq. (1) to fit the resulting expression to the experimental curve. As a result $r_{tip}$ and $tan(\gamma)$ are obtained as fitting parameters, as done in Fig. 4(c):

$$\frac{1}{2}w_{exp} = r_{tip} + (h - r_{tip})\tan(\gamma) + \frac{1}{2}h \tag{3}$$

Eventually, the reader could prefer to develop Eq. (3) to obtain $r_{tip}$ and $tan(\gamma)$ from a linear regression as follow:

$$w_{exp} = 2r_{tip} + 2(h - r_{tip})\tan(\gamma) + h \tag{4}$$
$$w_{exp} = 2r_{tip} + 2h\tan(\gamma) - 2r_{tip}\tan(\gamma) + h \tag{5}$$
$$w_{exp} = 2r_{tip} * [1 - \tan(\gamma)] + h * [2\tan(\gamma) + 1] \tag{6}$$

So, the dependence of $h$ on $w_{exp}$

$$h = \frac{w_{exp}}{[2\tan(\gamma)+1]} - \frac{2r_{tip*}(1-\tan(\gamma))}{[2\tan(\gamma)+1]} \tag{7}$$

Where the slope is $= \frac{1}{[2\tan(\gamma)+1]} \tag{9}$

Using the data from Fig. 4 we would find $h = 0{,}5924w_{exp}$ - $11{,}703$ nm which leads to $\gamma = 19.0$ degrees and $r_{tip} = 15.1$ nm. As discussed in the manuscript the determination of $r_{tip}$ by means of h-w method is affected by the error in the determination of $tan(\gamma)$, see Eq. (9).

## SI5.2 Tip radius larger than the object height.

In general, the object size will be shorter than the tip radius when using metal coated probes, such as conductive or MFM tips. Of course, we can use analogous method, however, the expressions would be rather different because in this operation range:

$$\frac{1}{2} w_{exp} = \Delta + \frac{1}{2} w \quad (9)$$

$$\Delta = r_{tip} * Cos\left[ArcSin\left(\frac{r_{tip}-h}{r_{tip}}\right)\right] \quad (10)$$

Developing Eq. (10) by squares we have that:

$$\Delta^2 = r_{tip}^2 * \left[1 - \left(\frac{r_{tip}-h}{r_{tip}}\right)^2\right] \quad (11)$$

$$\Delta^2 = r_{tip}^2 - r_{tip}^2 \left(\frac{r_{tip}^2 + h^2 - 2r_{tip}*h}{r_{tip}^2}\right) \quad (12)$$

$$\Delta^2 = r_{tip}^2 - r_{tip}^2 - h^2 + 2r_{tip} * h \quad (13)$$

$$\Delta^2 = 2r_{tip} * h - h^2 \quad (14)$$

Considering *w=h* (as applies for cubic objects) and introducing Eq. (14) in Eq. (9),

$$w_{exp} = h + 2\sqrt{2h\, r_{tip} - h^2} \quad (15)$$

In this case, we can plot w-h to determine the tip as a fitting parameter to Eq. (15). Obviously, we cannot get information about the tip-to-face angle in this operation range. In Fig. SI5 we show the w-h curve for a MFM tip. The corresponding tip reconstruction is carried out in Fig. SI6.

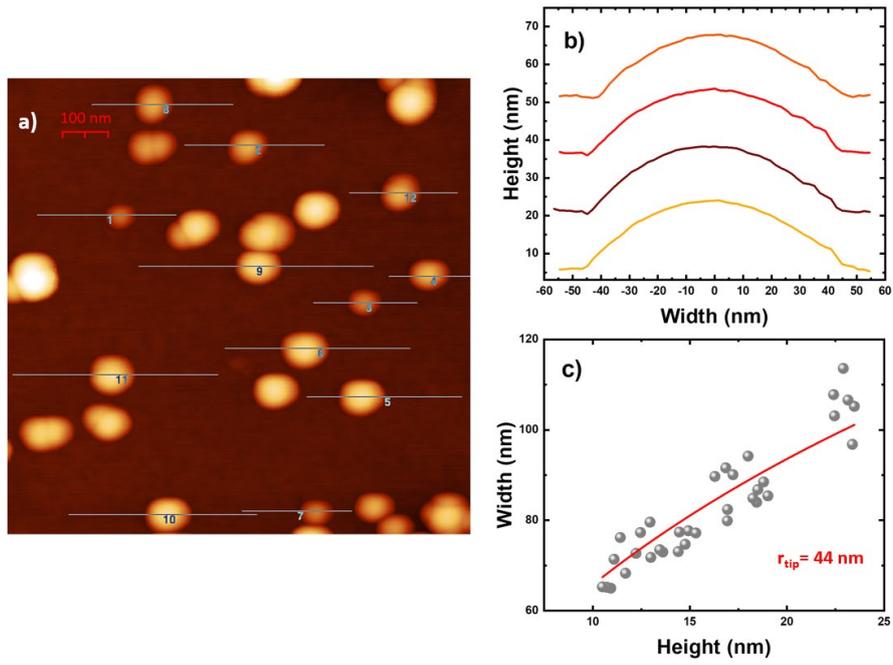

**Figure SI5.** W-h curve for a MFM tip, analogous to Fig. 4 found in the manuscript.

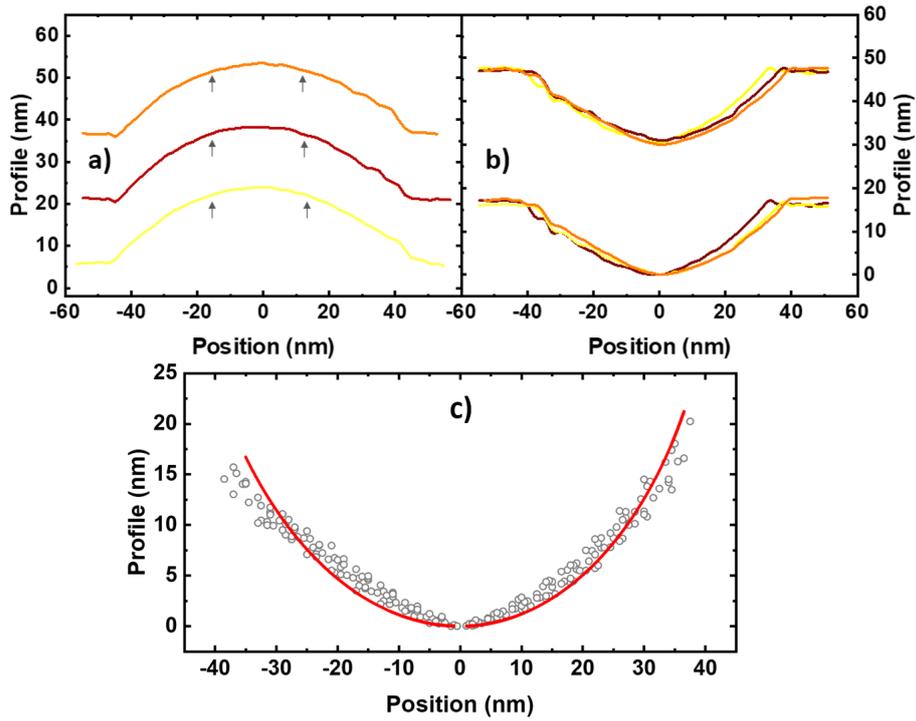

**Figure SI6.** Tip reconstruction of a MFM tip, analogous to Fig. 5 found in the manuscript.